# A Comparative Study of Load Balancing Algorithms in Cloud Computing Environment

Mayanka Katyal*, Atul Mishra**

**Abstract**

Cloud Computing is a new trend emerging in IT environment with huge requirements of infrastructure and resources. Load Balancing is an important aspect of cloud computing environment. Efficient load balancing scheme ensures efficient resource utilization by provisioning of resources to cloud user's on-demand basis in pay-as-you-say-manner. Load Balancing may even support prioritizing users by applying appropriate scheduling criteria. This paper presents various load balancing schemes in different cloud environment based on requirements specified in Service Level Agreement (SLA).

## 1. INTRODUCTION

Cloud Computing is made up by aggregating two terms in the field of technology. First term is Cloud and the second term is computing. Cloud is a pool of heterogeneous resources. It is a mesh of huge infrastructure and has no relevance with its name "Cloud". Infrastructure refers to both the applications delivered to end users as services over the Internet and the hardware and system software in datacenters that is responsible for providing those services. In order to make efficient use of these resources and ensure their availability to the end users "Computing" is done based on certain criteria specified in SLA. Infrastructure in the Cloud is made available to the user's On-Demand basis in pay-as-you-say-manner.

Computation in cloud is done with the aim to achieve maximum resource utilization with higher availability at minimized cost.

### 1.1. Cloud v/s Cluster and Grid

Clusters [1] are parallel and distributed systems, governed under the supervision of single administrative domain. The node (stand-alone computers) in the cluster integrates to form a single computing resource.

Grid [1] is aggregation of autonomous resources that are geographically distributed. The nodes in grid permit sharing and selection dynamically at runtime.

Clouds [1], [2] are not the combination of clusters and grid but are next generation to clusters and grid. Similar to Cluster and Grid, Cloud is also a collection of parallel and distributed systems. Cloud is not a single domain. Unlike cluster and grid, cloud has multiple domains and the nodes of cloud are "Virtualized" [9].

### 1.2. Cloud Perspectives

Cloud has different meaning to different stakeholders. There are three main stakeholders of cloud:

### 1.2.1. End Users

These are the customers or consumers of cloud. They use the various services (Infrastructure/ Software/Platform) provided by the cloud. Before using the cloud services, the users of cloud must agree to the Service Level Agreement (SLA) specified by the Cloud Provider.

___________

\* Student, Department of Computer Engineering, YMCA University of Science and Technology, Faridabad, Haryana, India. E-mail: mayankakatyal@gmail.com

\*\* Associate Professor, Department of Computer Engineering, YMCA University of Science and Technology, Faridabad, Haryana, India. E-mail: mish.atul@gmail.com

They use the services on demand basis and have to pay for the services availed depending upon their usage. Cloud provides its users flexibility in availing its services by incorporating utility computing.

Prior to signing of SLA, the users of cloud must verify that SLA contains certain Quality-of-Service (QoS) parameters which are pre-requisites of the consumer, before using cloud services. Some of the basic requirements or issues of cloud users is listed in Table 1.

Hence, for end user, Cloud Computing is a scenario where the user can have access to any kind of infrastructure, software or platform in a secure manner- at reduced cost- on demand basis-in an easy to use manner.

### 1.2.2. Cloud Provider

Cloud provider can offer either public or private or hybrid cloud. They are responsible for building of the cloud.

Private clouds [4] are owned by enterprises or business for their internal use. They may use it to store and manage Big-Data of their organization or to provide enough resources on demand basis to its team of employees or clients. They offer greatest level of security.

OpenStack [5], VMware [6] and CloudStack [7] are private clouds.

Public clouds [4] may be used by individuals or an organization based upon their requirements and necessities. They offer greatest level of efficiency in shared resources. Confidentiality is the major security issue in using public cloud. They are more vulnerable than private clouds. Amazon web services [8], Google Compute Engine [9], Microsoft Azure [10], HP cloud [11] are some of the public clouds.

A hybrid cloud [4] is a combination of public and private cloud. It allows businesses to manage some resources internally within organization and some externally. The downside is that the complexity of overall management increases along with security concerns. To optimize the use of one or more combination of private or public clouds CliQr [11] allows the businesses to accommodate changing needs of users.

Cloud provider must accomplish its job of "resource provisioning". Resource provisioning includes two main tasks. These include managing of huge bundle of resources that make up cloud and providing these resources to the end users. Several provisioning related issues are mentioned in Table 1.

**Table 1:** Stakeholders of Cloud

| Type of stakeholder | Requirements/Issues |
|---|---|
| End User | Security<br>Provenance<br>Privacy<br>High Availability<br>Reduced Cost<br>Ease-of-use |
| Cloud Provider | Managing Resources<br>Outsourcing<br>Resource Utilization<br>Energy Efficiency<br>Metering<br>Providing Resources<br>Cost Efficiency<br>Meet end user requirements<br>Utility Computing |
| Cloud Developer | Elasticity/ Scalability<br>Virtualization<br>Agility and Adaptability<br>Availability<br>Data Management<br>Reliability<br>Programmability |

### 1.2.3. Cloud Developer

This entity lies between end user and cloud provider. Cloud developer has the responsibility of taking into consideration both the perspectives of the cloud (i.e. view of end user and cloud provider). The developer of cloud must adhere to all the technical details of the cloud which are essential to meet the requirements of both, the cloud user as well as the cloud provider. Some of the basic issues that cloud developer must focus on are given in Table 1. Main motive of the developer is to bridge the gap between the end user of the cloud and the cloud provider.

This paper discusses different Load Balancing schemes in cloud computing environment. The rest of the paper is organized as follows: Section 2 discusses load balancing in cloud computing environment. Section 3 discusses some related work to load balancing algorithms. We conclude our study in Section 4.

## 2. LOAD BALANCING IN CLOUD COMPUTING ENVIRONMENT

Load balancing in cloud computing provides an efficient solution to various issues residing in cloud computing environment set-up and usage. Load balancing must take into account two major tasks, one is the resource provisioning or resource allocation and other is task scheduling in distributed environment. Efficient provisioning of resources and scheduling of resources as well as tasks will ensure:

a. Resources are easily available on demand.
b. Resources are efficiently utilized under condition of high/low load.
c. Energy is saved in case of low load (i.e. when usage of cloud resources is below certain threshold).
d. Cost of using resources is reduced.

For measuring the efficiency and effectiveness of Load Balancing algorithms simulation environment are required. CloudSim [12] is the most efficient tool that can be used for modeling of Cloud. During the lifecycle of a Cloud, CloudSim allows VMs to be managed by hosts which in turn are managed by datacenters.

Cloudsim provides architecture with four basic entities. These entities allow user to set-up a basic cloud computing environment and measure the effectiveness of Load Balancing algorithms. A typical Cloud modeled using CloudSim consists of following four entities Datacenters, Hosts, Virtual Machines and Application as well as System Software. Datacenters entity has the responsibility of providing Infrastructure level Services to the Cloud Users. They act as a home to several Host Entities or several instances hosts' entities aggregate to form a single Datacenter entity. Hosts in Cloud are Physical Servers that have pre-configured processing capabilities. Host is responsible for providing Software level service to the Cloud Users. Hosts have their own storage and memory. Processing capabilities of hosts is expressed in MIPS (million instructions per second). They act as a home to Virtual Machines or several instances of Virtual machine entity aggregate to form a Host entity. Virtual Machine allows development as well as deployment of custom application service models. They are mapped to a host that matches their critical characteristics like storage, processing, memory, software and availability requirements. Thus, similar instances of Virtual Machine are mapped to same instance of a Host based upon its availability. Application and System software are executed on Virtual Machine on-demand.

Class diagram of Cloud architecture illustrating relationship between the four basic entities is shown in fig 1. Thus, the object oriented approach of CloudSim can be used to simulate Cloud Computing environment.

**Figure 1.** Class Diagram of Cloud

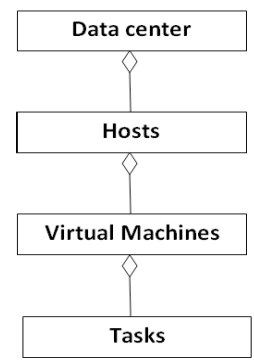

### 2.1. Resource Allocation

Resource provisioning is the task of mapping of the resources to different entities of cloud on demand basis. Resources must be allocated in such a manner that no node in the cloud is overloaded and all the available resources in the cloud do not undergo any kind of wastage (wastage of bandwidth or processing core or memory etc.). Mapping of resources to cloud entities is done at two levels:

### 2.1.1. VM Mapping onto the Host

Virtual machines reside on the host (physical servers). More than one instance of VM can be mapped onto a single host subject to its availability and capabilities. Host is responsible for assigning processing cores to VM. Provisioning policy define the basis of allocating processing cores to VM on demand. Allocation policy or algorithm must ensure that critical characteristics of Host and VM do not mismatch.

### 2.1.2 Application or Task Mapping onto VM

Applications or tasks are actually executed on VM. Each application requires certain amount of processing power for their completion. VM must provide required processing power to the tasks mapped onto it. Tasks

**Table 2:** Comparison between Resource Allocation and Task Scheduling

| Task | Sub-Category | Issues Resolved | Provider Oriented | Customer Oriented |
|---|---|---|---|---|
| Resource Allocation | At host level<br>At VM level | Efficient Utilization<br>Minimize Makespan<br>Ensure Availability | Yes | Yes |
| Task Scheduling | Space-Sharing<br>Time-Sharing | Minimize overall response time | No | Yes |

must be mapped onto appropriate VM based upon its configuration and availability.

## 2.2. Task Scheduling

Task scheduling is done after the resources are allocated to all cloud entities. Scheduling defines the manner in which different entities are provisioned. Resource provisioning defines which resource will be available to meet user requirements whereas task scheduling defines the manner in which the allocated resource is available to the end user (i.e. whether the resource is fully available until task completion or is available on sharing basis). Task scheduling provides "Multiprogramming Capabilities" in cloud computing environment.

Task scheduling can be done in two modes:

   a. Space shared
   b. Time shared

Both hosts and VM can be provisioned to users either in space shared mode or time shared mode.

In space sharing mode resources are allocated until task does not undergo complete execution (i.e. resources are not preempted); whereas in time sharing mode resources are continuously preempted till task undergoes completion.

Table 2. gives the comparison of resource allocation and task scheduling and specifies the issues resolved by each technique of load balancing.

Based on resource provisioning and scheduling, four cases can be examined under different performance criteria so as to get efficient load balancing scheme.

Case 1: Hosts and VMs, both are provisioned in space sharing manner.

Case 2: Hosts and VMs, both are provisioned to VMs and tasks respectively in time sharing manner.

Case 3: Hosts are provisioned to VMs in space sharing manner and VMs are provisioned to tasks in time sharing manner.

Case 4: Hosts are provisioned to VMs in time sharing manner and VMs are provisioned to tasks in space sharing manner.

## 3. RELATED WORK TO LOAD BALANCING ALGORITHMS

Cloud is made up of massive resources. Management of these resources requires efficient planning and proper layout. While designing an algorithm for resource provisioning on cloud the developer must take into consideration different cloud scenarios and must be aware of the issues that are to be resolved by the proposed algorithm. Therefore, resource provisioning algorithm can be categorized into different classes based upon the environment, purpose and technique of proposed solution.

### 3.1. Load Balancing on the basis of Cloud Environment

Cloud computing can have either static or dynamic environment based upon how developer configures the cloud demanded by the cloud provider.

#### 3.1.1. Static Environment

In static environment the cloud provider installs homogeneous resources. Also the resources in the cloud are not flexible when environment is made static. In this scenario, the cloud requires prior knowledge of nodes capacity, processing power, memory, performance and statistics of user requirements. These user requirements are not subjected to any change at run-time. Algorithms proposed to achieve load balancing in static environment cannot adapt to the run time changes in load. Although static environment is easier to simulate but is not well suited for heterogeneous cloud environment.

Round Robin algorithm [13] provides load balancing in static environment. In this the resources are provisioned to the task on first-cum-first-serve (FCFS- i.e. the task that entered first will be first allocated the resource) basis and scheduled in time sharing manner. The resource which is least loaded (the node with least number of connections) is allocated to the task. Eucalyptus uses greedy (first-fit) with round-robin for VM mapping.

Radojevic proposed an improved algorithm over round robin called CLBDM (Central Load Balancing Decision Model) [14]. It uses the basis of round robin but it also measures the duration of connection between client and server by calculating overall execution time of task on given cloud resource.

### 3.1.2. Dynamic Environment

In dynamic environment the cloud provider installs heterogeneous resources. The resources are flexible in dynamic environment. In this scenario cloud cannot rely on the prior knowledge whereas it takes into account run-time statistics. The requirements of the users are granted flexibility (i.e. they may change at run-time). Algorithm proposed to achieve load balancing in dynamic environment can easily adapt to run time changes in load. Dynamic environment is difficult to be simulated but is highly adaptable with cloud computing environment.

Based on WLC [15] (weighted least connection) algorithm, Ren proposed a load balancing technique in dynamic environment called ESWLC. It allocates the resource with least weight to a task and takes into account node capabilities. Based on the weight and capabilities of the node, task is assigned to a node. LBMM (Load Balancing Min-Min) algorithm proposed in paper [16] uses three level frameworks for resource allocation in dynamic environment. It uses OLB (opportunistic load balancing) algorithm as its basis. Since cloud is massively scalable and autonomous, dynamic scheduling is better choice over static scheduling.

### 3.2. Load Balancing based on Spatial Distribution of Nodes

Nodes in the cloud are highly distributed. Hence the node that makes the provisioning decision also governs the category of algorithm to be used. There can be three types of algorithms that specify which node is responsible for balancing of load in cloud computing environment.

### 3.2.1. Centralized Load Balancing

In centralized load balancing technique all the allocation and scheduling decision are made by a single node. This node is responsible for storing knowledge base of entire cloud network and can apply static or dynamic approach for load balancing. This technique reduces the time required to analyze different cloud resources but creates a great overhead on the centralized node. Also the network is no longer fault tolerant in this scenario as failure intensity of the overloaded centralized node is high and recovery might not be easy in case of node failure.

### 3.2.2. Distributed Load Balancing

In distributed load balancing technique, no single node is responsible for making resource provisioning or task scheduling decision. There is no single domain responsible for monitoring the cloud network instead multiple domains monitor the network to make accurate load balancing decision. Every node in the network maintains local knowledge base to ensure efficient distribution of tasks in static environment and re-distribution in dynamic environment.

In distributed scenario, failure intensity of a node is not neglected. Hence, the system is fault tolerant and balanced as well as no single node is overloaded to make load balancing decision.

Comparison of different static and dynamic load balancing algorithms is given in Table 3. It also compares them on the basis of spatial distribution of nodes.

A nature inspired solution is presented in paper [17] called Honeybee Foraging for load balancing in distributed scenario. In Honeybee foraging the movement of ant in search of food forms the basis of distributed load balancing in cloud computing environment. This is a self organizing algorithm and uses queue data structure for its implementation. Biased random sampling [18] is another distributed load balancing technique which uses virtual graph as the knowledge base.

**Table 3:** Comparison Table of Load Balancing Algorithms in Cloud Computing Environment

| Algorithm | Static Environment | Dynamic Environment | Centralized Balancing | Distributed Balancing | Hierarchical Balancing |
|---|---|---|---|---|---|
| Round-robin | Yes | No | Yes | No | No |
| CLBDM[22] | Yes | No | Yes | No | No |
| Ant Colony[20] | No | Yes | No | Yes | No |
| Map Reduce[9] | Yes | No | No | Yes | Yes |
| Particle Swarm Optimization [21] | No | Yes | No | Yes | No |
| MaxMin[22] | Yes | No | Yes | No | No |
| MinMin[22] | Yes | No | Yes | No | No |
| Biased Random Sampling | No | Yes | No | Yes | No |
| Active Clustering[18] | No | Yes | No | Yes | No |
| LBMM | No | Yes | No | No | Yes |
| OLB[23] | Yes | No | Yes | No | No |
| WLC | No | Yes | Yes | No | No |
| ESWLC | No | Yes | Yes | No | No |
| Genetic Algorithm[24] | No | Yes | Yes | No | No |

### 3.2.3. Hierarchical Load Balancing

Hierarchical load balancing involves different levels of the cloud in load balancing decision. Such load balancing techniques mostly operate in master slave mode. These can be modeled using tree data structure wherein every node in the tree is balanced under the supervision of its parent node. Master or manager can use light weight agent process to get statistics of slave nodes or child nodes. Based upon the information gathered by the parent node provisioning or scheduling decision is made.

Three-phase hierarchical scheduling proposed in paper [19] has multiple phases of scheduling. Request monitor acts as a head of the network and is responsible for monitoring service manager which in turn monitor service nodes. First phase uses BTO (Best Task Order) scheduling, second phase uses EOLB (Enhanced Opportunistic Load Balancing) scheduling and third phase uses EMM (Enhanced Min-Min) scheduling.

### 3.3. Load Balancing Based on Task Dependencies

Dependent tasks are those whose execution is dependent on one or more sub-tasks. They can be executed only after completion of the sub-tasks on which it is dependent. Therefore, scheduling of such task prior to execution of sub-tasks is in-efficient. Task dependency is modeled using workflow based algorithms.

Workflow basically uses DAG [25] as knowledge base to represent task dependency. Different workflow based solution consider different parameters. Algorithm are designed keeping in mind whether single or multiple workflows are to be modeled or single or multiple QoS parameters are to be maintained in the system. Different workflows with or without completely different structure are termed as multiple workflows. Workflows can also be classified as Transaction Incentive (multiple instances of one workflow that have same structure) and Data Incentive workflows (size and quantity of data is large).

Cost based scheduling algorithm in [26] is designed for single workflows. It partitions the workflows and assigns each partition a deadline. Zhifeng Yu and Weisong Shi designed an algorithm for multiple workflows which focus only on execution time. With an aim of maximizing throughput Ke Lie et al proposed scheduling strategy which is meant for transaction incentive workflows. For clouds based on Hadoop CloudWF (computational workflow system) encodes workflow blocks and block-to-block dependencies. Hadoop HBase sparse table is used to store information related to workflows. It is fault tolerant and uses map-reduce framework.

**Table 4:** Comparison of different Types of Load Balancing Scenarios in Cloud Computing Environment

| Type of Algorithm | Knowledge Base | Issues to be addressed | Usage | Drawbacks |
|---|---|---|---|---|
| Static | Prior knowledge base is required about each node statistics and user requirements. | Response time<br>Resource utilization<br>Scalability<br>Power consumption and Energy Utilization<br>Makespan<br>Throughput/Performance | Used in homogeneous environment. | Not Flexible<br>Not scalable<br>Is not compatible with changing user requirements as well as load |
| Dynamic | Run time statistics of each node are monitored to adapt to changing load requirements. | Location of processor to which load is transferred by an overloaded processor.<br>Transfer of task to a remote machine.<br>Information Gathering.<br>Load estimation.<br>Limiting the number of migrations.<br>Throughput | Used in heterogeneous environment. | Complex<br>Time Consuming |
| Centralized | Single node or server is responsible for maintaining the statistics of entire network and updating it from time to time. | Threshold policies<br>Throughput<br>Failure Intensity<br>Communication between central server and processors in network.<br>Associated Overhead | Useful in small networks with low load. | Not fault tolerant<br>Overloaded central decision making node |
| Distributed | All the processors in the network responsible for load balancing store their own local database (e.g. MIB) to make efficient balancing decisions. | Selection of processor that take part in load balancing.<br>Migration time<br>Interprocessor communication<br>Information exchange criteria<br>Throughput<br>Fault tolerance | Useful in large and heterogeneous environment. | Algorithm complexity<br>Communication overhead |
| Hierarchical | Nodes at different levels of hierarchy communicate with the nodes below them to get information about the network performance. | Threshold policies<br>Information exchange criteria<br>Selection of nodes at different levels of network<br>Failure intensity<br>Performance<br>Migration time | Useful in medium or large size network with heterogeneous environment. | Less fault tolerant<br>Complex |

| Type of Algorithm | Knowledge Base | Issues to be addressed | Usage | Drawbacks |
|---|---|---|---|---|
| Workflow Dependent | DAG is used to model dependencies of task and can be used to make scheduling decision. | Type of workflow<br>Single workflow<br>Multiple workflow<br>Transaction incentive workflows<br>Data incentive workflows<br>Fault tolerance<br>Execution time<br>Makespan<br>Migration time | Used in modeling of task dependencies in any kind of environment (either homogeneous or heterogeneous.) | Difficult to model<br>Maintenance of knowledge base is complex.<br>Higher Complexity |

Table 4, compares different type of load balancing scenarios in cloud computing environment. It specifies the knowledge base, usage and drawbacks of each type of algorithm and issues addressed by these algorithms.

## 4. CONCLUSION

Load Balancing is an essential task in Cloud Computing environment to achieve maximum utilization of resources. In this paper, we discussed various load balancing schemes, each having some pros and cons. On one hand static load balancing scheme provide easiest simulation and monitoring of environment but fail to model heterogeneous nature of cloud. On the other hand, dynamic load balancing algorithm are difficult to simulate but are best suited in heterogeneous environment of cloud computing. Also the level at node which implements this static and dynamic algorithm plays a vital role in deciding the effectiveness of algorithm. Unlike centralized algorithm, distributed nature of algorithm provides better fault tolerance but requires higher degree of replication and on the other hand, hierarchical algorithm divide the load at different levels of hierarchy with upper level nodes requesting for services of lower level nodes in balanced manner. Hence, dynamic load balancing techniques in distributed or hierarchical environment provide better performance. However, performance of the cloud computing environment can be further maximized if dependencies between tasks are modeled using workflows.